\documentclass[aps,showpacs,nofootinbib,superscriptaddress,twocolumn]{revtex4}

\usepackage{graphicx}
\usepackage{bm}
\usepackage{amsmath}
\usepackage{amssymb}
\usepackage{hyperref}

\begin{document}

\title{Predicting the $\sin\phi_S$ Transverse Single-spin Asymmetry of Pion Production at an Electron Ion Collider}
\author{Xiaoyu Wang}
\affiliation{Department of Physics, Southeast University, Nanjing 211189, China}
\author{Zhun Lu}
\email{zhunlu@seu.edu.cn}
\affiliation{Department of Physics, Southeast University, Nanjing 211189, China}

\begin{abstract}
  We study the transverse single-spin asymmetry with a $\sin\phi_S$ modulation in semi-inclusive deep inelastic scattering.
  Particularly, we consider the case in which the transverse momentum of the final state hadron is integrated out.
  Thus, the asymmetry is merely contributed by the coupling of the transversity distribution function $h_1(x)$ and the twist-3 collinear fragmentation function $\tilde{H}(z)$.
  Using the available parametrization of $h_1(x)$ from SIDIS data and the recent extracted result for $\tilde{H}(z)$,
  we predict the $\sin\phi_S$ asymmetry for charged and neutral pion production at an Electron Ion Collider.
  We find that the asymmetry is sizable and could be measured.
  We also study the impact of the leading-order QCD evolution effect and find that it affects the $\sin\phi_S$ asymmetry at EIC considerably.
\end{abstract}

\pacs{13.60.-r, 13.60.Le, 13.88.+e}
\maketitle

\section{Introduction}

Transverse single-spin asymmetry (SSA) in semi-inclusive processes has been recognized as a useful tool to probe the spin structure of nucleon.
One of the fundamental observables encoding the nucleon structure is the transversity distribution of quarks, denoted by $h_1^q(x)$, which describes the transverse polarization of the quark inside a transversely polarized nucleon: $h_1^q(x)=f_{q^\uparrow/p^\Uparrow}- f_{q^\downarrow/p^\Uparrow} $.
As a chiral-odd distribution, transversity is rather difficult to be probed, compared to the other two lead-twist collinear distribution functions: the unpolarized distribution and the helicity distribution, which have been extensively studied and measured.
To manifest the effect of transversity in a process, another chiral-odd function is needed to couple with transversity to ensure chirality conservation.
It has been shown that two promising processes may be applied to measure the transversity distributions.
One is the semi-inclusive deeply inelastic scattering (SIDIS)~\cite{Collins:1992kk}, and the other is the Drell-Yan process~\cite{Jaffe:1991ra}.

Under the framework of the transverse momentum dependent (TMD) factorization~\cite{Ji:2004wu}, in SIDIS the chiral-odd probe can be the Collins function $H_1^\perp(z,\bm{p}_{T}^2)$~\cite{Collins:1992kk}, a TMD fragmentation function that describes the fragmenting of a transversely polarized quark to an unpolarized hadron.
The corresponding observable is the Collins asymmetry with a $\sin(\phi_h+\phi_S)$ modulation.
Here $\phi_S$ and $\phi_h$ are the azimuthal angles of the transverse spin of the nucleon and the outgoing hadron, respectively.
The Collins asymmetry has been measured by the HERMES Collaboration~\cite{Airapetian:2010ds}, the COMPASS Collaboration~\cite{Alekseev:2010rw}, and the Jefferson Lab Hall A Collaboration~\cite{Qian:2011py}.
The SIDIS data combined with the $e^+\,e^-$ annihilation data were applied to extract~\cite{transversitypara2,parameter} the transversity of the up and down quarks.
Another probe in SIDIS is the chiral-odd dihadron fragmentation function, which will give rise to a $\sin(\phi_R+\phi_S)$ asymmetry based on the collinear factorization formalism.
In this approach two hadrons fragmented from a projectile quark should be detected.
The idea was used to extract the transversity in Ref.~\cite{Bacchetta:2011ip}.
In the Drell-Yan process, the transversity may be accessed by measuring the double transverse spin asymmetry, which is contributed by the convolution of the quark transveresity and the antiquark transversity.

In this work, we propose an alternative approach to access transversity in SIDIS.
We will show that the twist-3 chiral-odd fragmentation function, denoted by $\tilde{H}(z)$, can serve as a ``spin analyzer" to probe the transversity distribution.
As demonstrated in Ref.~\cite{0611265semi},  within the collinear picture (or equivalently, the transverse momentum of the outgoing hadron is integrated out), only the coupling between the function $\tilde{H}(z)$ and the transversity remains as a contribution to the $\sin\phi_S$ azimuthal modulation in the leptoproduction of single hadron off a transversely polarized nucleon.
The advantage of this approach is that the transverse momentum of the final-state hadron is not necessarily to be measured, in contrast to the Collins effect.
However, the feasibility of this approach has never been tested so far, partly because of the very limited knowledge of the almost unknown function $\tilde{H}(z)$.
This situation may be changed, since recent phenomenological studies~\cite{Metz:2012ct,1404.1033} of the transverse SSA in $p p^\uparrow \rightarrow \pi+ X$ provide very useful constraints on $\tilde{H}(z)$.
Particularly, the authors in Ref.~\cite{1404.1033} have adopted the SSA data at STAR~\cite{Adams:2003fx,Abelev:2008af,Adamczyk:2012xd,HEPPELMANN:2013ewa} and BRAHMS~\cite{Lee:2007zzh} of RHIC to extract the function $\hat{H}_{FU}^{\Im}(z,z_1)$ within the framework of collinear twist-3 factorization.
The function $\hat{H}_{FU}^{\Im}(z,z_1)$ is the imaginary part of the twist-3 fragmentation function $H_{FU}(z,z_1)$, which involves the F-type quark-quark-gluon correlation~\cite{Yuan:2009dw,Kang:2010zzb,Metz:2012ct}; and it can be connected to $\tilde{H}(z)$ by the integral
\begin{equation}
\tilde H^{h/q}(z)=2z^3\int_z^\infty\frac{dz_1}{z_1^2}\frac{1}{\frac{1}{z}-\frac{1}{z_1}}
\hat{H}^{h/q,\Im}_{FU}(z,z_1)\,.\label{eq:HZHFU}
\end{equation}
A calculation using the spectator model also shows that the size of $\tilde{H}(z)$ is substantial~\cite{Lu:2015wja}.
Therefore, it is intriguing to study the consequence of sizable $\tilde{H}(z)$ in the $\sin\phi_S$ asymmetry in SIDIS.
More over, the study of the asymmetry in SIDIS may provide a check whether $\tilde{H}(z)$ can cause the target SSAs in proton-proton collisions.
Since the SSA data used to extract $\hat{H}_{FU}^{\Im}(z,z_1)$ are collected in the rather high energy region (mainly $\sqrt{s}=200$ GeV) for which typically $P_{h\perp}> 1\mathrm{GeV}$, where the collinear twist-3 factorization is applicable, we present the prediction of the $\sin{\phi_S}$ asymmetry in SIDIS at a future Electron Ion Collider (EIC)~\cite{Accardi:2012qut}.
We calculate the $\sin{\phi_S}$ asymmetry of charged and neutral pion production at the energy $\sqrt{s}=45$ GeV, which is comparable to the RHIC energy.
To do this, we adopt the available parametrization of the transversity for up and down quarks~\cite{parameter} and the recent extraction for $\hat{H}^{\pi^+/u,\bar{d},\Im}_{FU}(z,z_1)$~\cite{1404.1033}.
Furthermore, we include the leading-order (LO) QCD evolution of the transversity distribution and $\tilde{H}(z)$, and compare the corresponding results to those without evolution.

The remained content of the paper is organized as follows.
In Section. II, we set up the formalism of the $\sin\phi_S$ asymmetry in SIDIS in the collinear picture.
In Section. III, we present the numerical calculation of the asymmetries in the leptoproduction of charged and neutral pions at EIC.
Some conclusion is addressed in Sec.~\ref{conclusion}.

\section{Formalism of the $\sin\phi_S$ Asymmetry in SIDIS}

\label{sect.aut_non}

\begin{figure*}
\centering
\includegraphics[width=0.99\columnwidth]{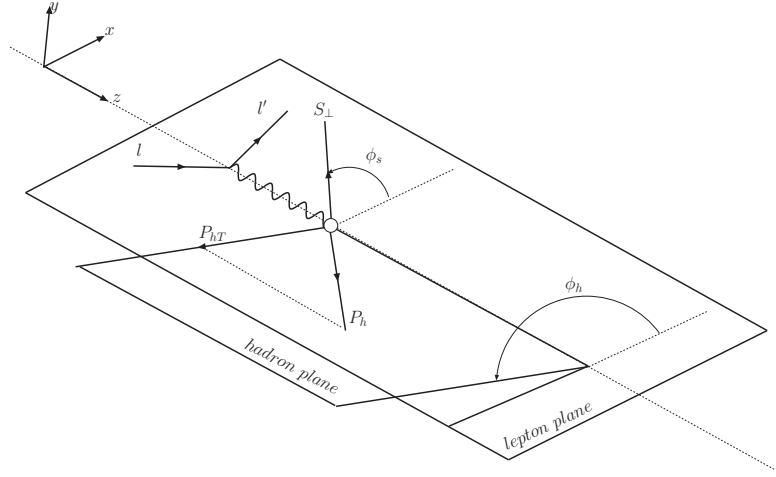}
\caption{The definition of the azimuthal angles in SIDIS. $P_{h}$ stands for the momentum of the produced hadron, $S_\perp$ is the transverse component of the spin vector $S$ with respect to the virtual photon momentum.}
\label{lepton-hadron plane}
\end{figure*}

The process we study is the pion electroproduction off a transversely polarized proton target:
\begin{equation}
\label{eq:sidis}
e(\ell)+p^\uparrow(P) \longrightarrow e(\ell^\prime)+{\pi}(P_h)+X(P_X),
\end{equation}
where $l$ and $l'$ stand for the momenta of the incoming and outgoing leptons, and $P$ and $P_h$ denote the momenta of the target nucleon and the final-state hadron (in our case the hadron is the pion meson), respectively.
The reference frame of the process under study is shown in Fig.~\ref{lepton-hadron plane}, in which the momentum of virtual photon defines the axis of $z$, $\phi_h$ denotes the the azimuthal angle of the final hadron around the virtual photon, and $\phi_S$ stands for the angle between the lepton scattering plane and the direction of the transverse spin of the nucleon target.

The invariants used to express the differential cross section are defined as
\begin{align}
x=\frac {Q^2}{2P \cdot q}\,,\quad y=\frac{P \cdot q}{P \cdot l}\,,\quad z=\frac{P \cdot P_h}{P \cdot q}\,,\nonumber \\
 \gamma=\frac{2Mx}{Q}\,,\quad Q^2=-q^2\,,\quad s=(P+l)^2\,. \nonumber
\end{align}
As usual, $q=\ell-\ell^\prime$ is defined as the momentum of the virtual photon, and $M$ denotes the proton mass.
Up to twist-3 level, the six-fold ($x$, $y$, $z$, $\phi_h$, $\phi_S$ and $P_{hT}$) differential cross section in SIDIS with a transversely polarized target has the general form~\cite{0611265semi}:
\begin{align}
&\frac{d^6\sigma}{dxdydzd\phi_h d\phi_S dP_{hT}^2}=\frac{\alpha^2}{xyQ^2}\frac{y^2}{2(1-\varepsilon)}(1+\frac{\gamma^2}{2x})\nonumber\\
&\times \sqrt{2\varepsilon(1+\varepsilon)}\left\{\sin \phi_S \,F^{\sin\phi_S}_{UT}(x,z,P_T)\right.\nonumber\\
&+\sin (2\phi_h-\phi_S) \,F^{\sin(2\phi_h-\phi_S)}_{UT}(x,z,P_T) \nonumber\\
& + \left. \textrm{leading twist terms}\right\}\,, \label{eq:diffcs1}
\end{align}
where $\varepsilon$ is the ratio of the longitudinal and transverse photon flux
\begin{equation}
\label{eq:epsilon}
\varepsilon=\frac{1-y-\frac{1}{4}\gamma^2y^2}{1-y+\frac{1}{2}y^2+\frac{1}{4}\gamma^2y^2}.
\end{equation}

In Eq.~(\ref{eq:diffcs1}), $F^{\sin\phi_S}_{UT}$ and $F^{\sin(2\phi_h-\phi_S)}_{UT}$ are the twist-3 structure functions which contribute to the $\sin\phi_S$ and the $\sin (2\phi_h-\phi_S)$ azimuthal asymmetries, respectively.
Particularly,  $F^{\sin\phi_S}_{UT}(x,z,P_{hT})$ can be expressed as~\cite{0611265semi}
\begin{align}
F_{UT}^{\sin\phi_S}(x,z,P_{hT})&=\frac{2M}{Q}{\mathcal{C}}\left\{\left(x f_T D_1 - \frac{M_h}{M} h_1\frac{\tilde{H}}{z}\right)\right.\nonumber\\
&-\frac{\bm k_T \cdot \bm p_T}{2M M_h}\left[\left(x h_T H_1^\perp+\frac{M_h}{M}g_{1T}\frac{\tilde{G^\perp}}{z}\right)\right.\nonumber\\
&\left.\left.-\left(x h^\perp_T H_1^\perp-\frac{M_h}{M}f_{1T}^\perp \frac{\tilde{D^\perp}}{z}\right)\right]\right\}\,,\label{eq:fut1}
\end{align}
where $k_T$ and $p_T$ are the transverse momenta of the incoming and outgoing quarks, $M_h$ is the mass of the outgoing hadron, and the notation $\mathcal{C}[\omega f D]$ defines the convolution:
\begin{align}
\mathcal{C}[\omega f D]&=x\sum_q e_q^2\int d^2\bm{p}_T d^2\bm{k}_T\delta^{(2)}\left(\bm p_T-\bm k_T-\bm P_{hT}/z\right)\,\nonumber \\
&\omega(\bm p_T,\bm k_T)f^q(x,\bm{k}_T^2)D^q(z,\bm{p}_{T}^2).
\end{align}
Eq.~(\ref{eq:fut1}) contains convolutions of the twist-3 distributions and the twist-2 fragmentation functions, as well as convolutions of the twist-3 fragmentation functions and the twist-2 distribution functions.

In Ref.~\cite{Mao:2014aoa}, the role of the twist-3 TMD distributions $h_T(x,\bm{k}_T^2)$ and  $h^\perp_T(x,\bm{k}_T^2)$ in the $\sin\phi_S$ asymmetries as functions of $x$, $z$ and $P_{hT} = |\bm P_{hT}|$ was studied.
In this work, however, we will consider the particular case in which the transverse momentum of the outgoing pion meson is integrated out, or equivalently, the case in which only the longitudinal momentum fraction $z$ of pion is measured.
Thus, after $\int d^2\bm P_{hT}$ is performed, the differential cross section in Eq.~(\ref{eq:diffcs1}) turns to the form
\begin{align}
&\frac{d^4\sigma}{dxdydzd\phi_S}=\frac{2\alpha^2}{xyQ^2}\,\frac{y^2}{2(1-\varepsilon)}\,\left(1+\frac{\gamma^2}{2x}\right)\nonumber\\
&\times \,\sqrt{2\varepsilon(1+\varepsilon)}\,\sin \phi_S\, F^{\sin\phi_S}_{UT}\left(x,z\right)\,. \label{eq:diffcs2}
\end{align}
Here, the structure function $F^{\sin\phi_S}_{UT}\left(x,z\right)$ is the collinear counterpart of the original structure function $F^{\sin\phi_S}_{UT}\left(x,z,P_{hT}\right)$~\cite{0611265semi}
\begin{align}
&F^{\sin\phi_S}_{UT}\left(x,z\right)  = \int d^2 \bm{P}_{hT} F^{\sin\phi_S}_{UT}\left(x,z,P_{hT}\right)\nonumber \\
&=- \int d^2 \bm{P}_{hT} \frac{2M_h}{Q}\mathcal{C}\left [h_1\frac{\tilde H}{z}\right]\nonumber \\
&= -x\frac{2M_h}{Q} \sum_q e_q^2 \int d^2 \bm{k}_T\, z^2 \int d^2 \bm{p}_T h_1^q(x,\bm{k}_T^2)\frac{\tilde H^q(z,\bm{p}_T^2)}{z}\nonumber \\
& =  -x\sum_q e_q^2\frac{2M_h}{Q}h_1^q(x)\frac{\tilde H^q(z)}{z}\,,\label{eq:futcol}
\end{align}
where only the convolution of the transversity and the twist-3 collinear fragmentation function $\tilde H^q(z)$ remains.
In Eq.~(\ref{eq:futcol}) we have used the integration
\begin{equation}
\int d^2 \bm{P}_{hT} \,\delta^{(2)}\left(\bm p_T-\bm k_T-\bm P_{hT}/z\right) = z^2,
\end{equation}
and the Collinear fragmentation function is connected to the TMD function by
\begin{equation}
\tilde H^q(z) = z^2 \int d^2 \bm{p}_T \tilde H^q(z,\bm{p}_T^2).
\end{equation}
Therefore, the experimental investigation of the $\sin \phi_S $ azimuthal asymmetry without detecting $\bm P_{hT}$ may provide an alternative opportunity to measure the transversity.
To test the feasibility of this approach, we define the $x$-dependent $\sin \phi_S$ asymmetry
\begin{align}
\label{eq:autx}
&A_{UT}^{\sin\phi_S}(x) \nonumber \\
=&\frac{\int dy\int dz\frac{\alpha^2}{xyQ^2}\frac{y^2}{2(1-\epsilon)}(1+\frac{\gamma^2}{2x})\sqrt{2\epsilon(1+\epsilon)}F^{\sin\phi_s}_{UT}(x,z)}
{\int dy\int dz\frac{\alpha^2}{xyQ^2}\frac{y^2}{2(1-\epsilon)}(1+\frac{\gamma^2}{2x})F_{UU}(x,z)}\,,
\end{align}
where $F_{UU}$ is the unpolarized structure function:
\begin{equation}
\label{eq:fuu}
F_{UU}(x,z)=x\sum_q e_q^2f_1^q(x)D_1^q(z)\,,
\end{equation}
with $f_1^q(x)$ and $D_1^q(z)$ are the unpolarized distribution function and fragmentation function, respectively.
In a similar way, the $\sin \phi_S$ asymmetry as a function of $z$ can be written as
\begin{align}
\label{eq:autz}
&A_{UT}^{\sin\phi_s}(z) \nonumber\\
&=\frac{\int dx\int dy\frac{\alpha^2}{xyQ^2}\frac{y^2}{2(1-\epsilon)}(1+\frac{\gamma^2}{2x})
\sqrt{2\epsilon(1+\epsilon)}F^{\sin\phi_s}_{UT}(x,z)}
{\int dx\int dy\frac{\alpha^2}{xyQ^2}\frac{y^2}{2(1-\epsilon)}(1+\frac{\gamma^2}{2x})F_{UU}(x,z)}.
\end{align}

\begin{figure*}
\centering
\includegraphics[width=1.5\columnwidth]{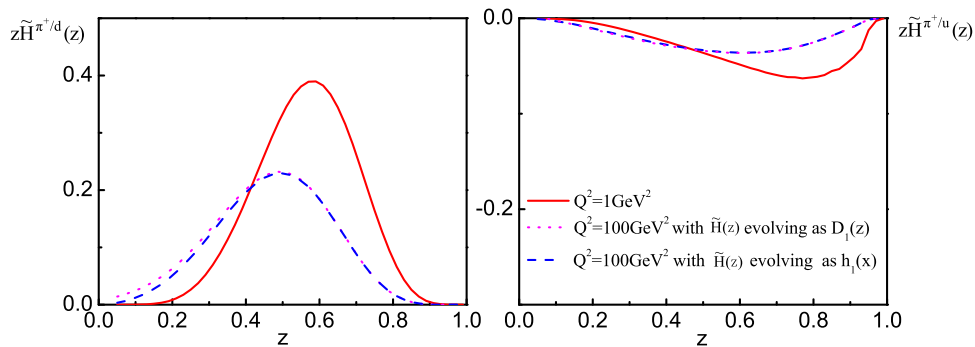}
\caption{Result of $z\tilde H^{\pi^+/d}(z)$ (left panel) and  $z\tilde H^{\pi^+/u}(z)$ (right panel) at the initial scale $Q^2\ =1\,\mathrm{GeV}^2$ (solid lines, taken from  Ref.~\cite{1404.1033}) and the evolved results at $Q^2\ =100\,\mathrm{GeV}^2$ (dotted lines: evolving as $D_1^q(z)$, dashed lines: evolving as $h_1^q(x)$; )}
\label{fig:htilde}
\end{figure*}

The function $\tilde H(z)$ appearing in Eq.~(\ref{eq:futcol}) is related to the imaginary part of the twist-3 fragmentation function $\hat{H}^{q}(z,z_1)$ via Eq.~(\ref{eq:HZHFU}).
Useful information for $\hat{H}^{q,\Im}_{FU}(z,z_1)$  may be obtained from Ref.~\cite{1404.1033}, in which the authors have extracted $\hat{H}^{(\pi/u,\bar{d}),\Im}_{FU}(z,z_1)$ from the data~\cite{Adams:2003fx,Abelev:2008af,Adamczyk:2012xd,HEPPELMANN:2013ewa,Lee:2007zzh} of $pp^\uparrow \rightarrow \pi X$ using the ansatz (at the scale $Q^2=1\,\textrm{GeV}^2$):
\begin{eqnarray}
\frac{\hat{H}_{FU}^{\pi^+/(u,\bar{d}),\,\Im}(z,z_{1})} {D_1^{\pi^+ / (u,\bar{d})}(z) \, D_1^{\pi^+ / (u,\bar{d})}(z/z_1)}
& = & \frac{N_{\textrm{fav}}}{2 I_{\textrm{fav}} J_{\textrm{fav}}} \, z^{\alpha_{\textrm{fav}}} (z/z_1)^{\alpha'_{\textrm{fav}}}
\nonumber \\
&& \hspace{-1.25cm} \times \, (1 - z)^{\beta_{\textrm{fav}}} \, (1 - z/z_{1})^{\beta'_{\textrm{fav}}} \,,
\label{eq:HFUIM}
\end{eqnarray}
with $N_{\textrm{fav}}$, $\alpha_{\textrm{fav}}$, $\alpha'_{\textrm{fav}}$, $\beta_{\textrm{fav}}$, $\beta'_{\textrm{fav}}$ the parameters in the model, and $I_{\textrm{fav}}$ and $J_{\textrm{fav}}$, the forms of which are given in Ref.~\cite{1404.1033}, are the functions of the above parameters.
The disfavored fragmentation functions $\hat{H}_{FU}^{\pi^+/(d,\bar{u}),\Im}$ are parameterized in full analogy to~(\ref{eq:HFUIM}) with the additional parameters $N_{\textrm{dis}}$, $\alpha_{\textrm{dis}}$, $\alpha'_{\textrm{dis}}$, $\beta_{\textrm{dis}}$, $\beta'_{\textrm{dis}}$.
The analysis in Ref.~\cite{1404.1033} shows that $\hat{H}^{q,\Im}_{FU}(z,z_1)$ plays an important role in the transverse SSA in the $p^\uparrow p \rightarrow \pi X$ process.
The $\pi^-$ fragmentation functions may be fixed through charge conjugation:
\begin{align}
\hat{H}^{\pi^-/(d, \bar{u}),\Im}_{FU}(z,z_1) =\hat{H}_{FU}^{\pi^+/(u,\bar{d}),\Im}(z,z_1)\, ,\\
\hat{H}^{\pi^-/(u, \bar{d}),\Im}_{FU}(z,z_1) =\hat{H}_{FU}^{\pi^+/(d,\bar{u}),\Im}(z,z_1)\, ,
\end{align}
and the $\pi^0$ fragmentation functions are given by the average of the fragmentation functions for $\pi^+$ and $\pi^-$.
Therefore, we will apply the parametrization for $\hat{H}^{h/q,\Im}$ in Ref.~\cite{1404.1033} and use Eq.~(\ref{eq:HZHFU}) to obtain $\tilde{H}^{q}(z)$ needed in the calculation.

For the transversity distribution function $h_1^q(x)$, we adopt the standard parameterization from Ref.~\cite{parameter} (at the initial scale $Q^2=2.41 \textrm{GeV}^2$)£º
\begin{equation}
\label{eq:transversity}
h_1^q(x)=\frac{1}{2}\mathcal{N}_q^T(x)[f_{1}^q(x)+g_1^q(x)]\,,
\end{equation}
with
\begin{equation}
\label{eq:n}
 \mathcal{N}_q^T(x)=N_q^T\,x^\alpha(1-\beta)^\beta\frac{(\alpha+\beta)^{\alpha+\beta}}{\alpha^\alpha\beta^\beta}\,.
\end{equation}
The values of the parameters $N_q^T, \alpha$, and $\beta$ in Eq. (\ref{eq:n}) are taken from Table. II in Ref.~\cite{parameter}.
In order to be in consistence with the choices in Ref.~\cite{parameter},  we apply the parametrization for the unpolarized distribution $f_1^q(x)$ from Ref.~\cite{9806404grv98} and that for the helicity distribution $g_1^q(x)$ from Ref.~\cite{helicity}, respectively.
For the unpolarized integrated fragmentation function $D_1^q(z)$, we adopt the leading-order (LO) set the DSS parametrization~\cite{0703242fdss}, which is also chosen in Ref.~\cite{1404.1033}.

\section{Numerical Results at an EIC}
\label{Sec.calculation}

\begin{figure*}
\centering
\includegraphics[width=1.7\columnwidth]{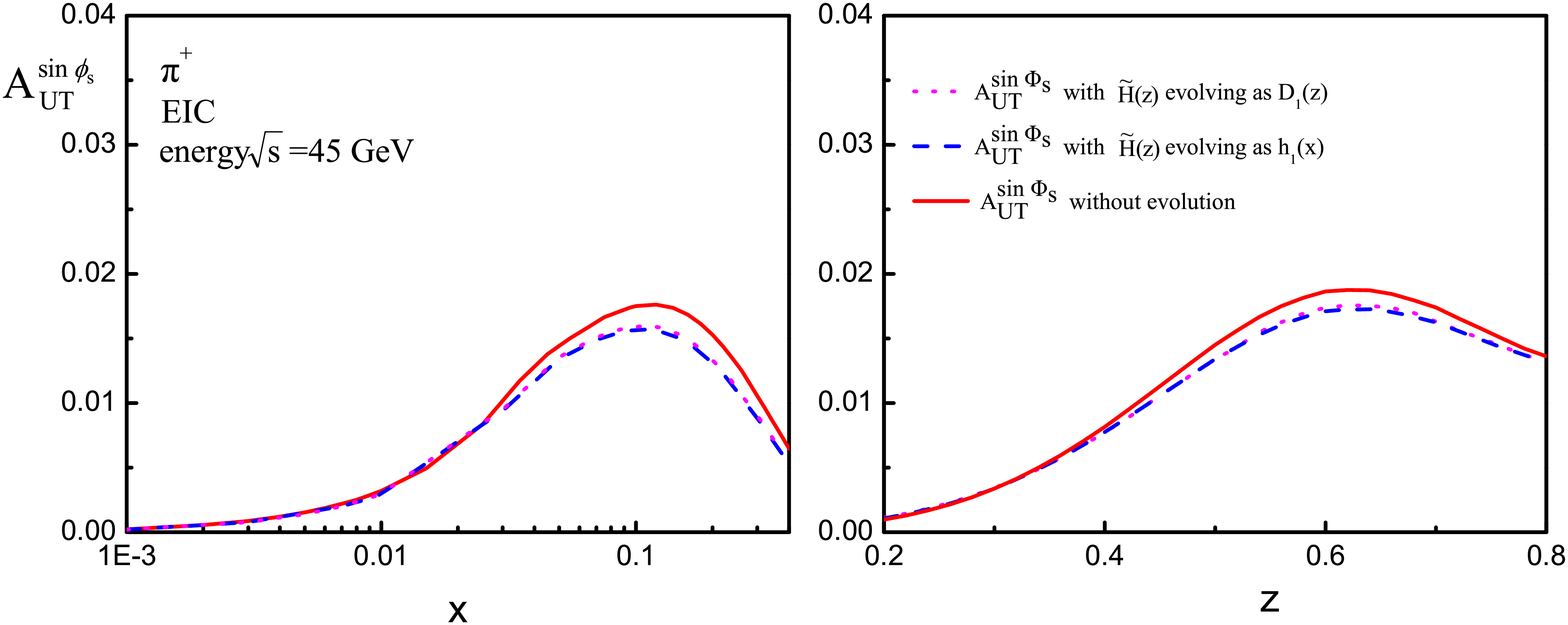}
\includegraphics[width=1.7\columnwidth]{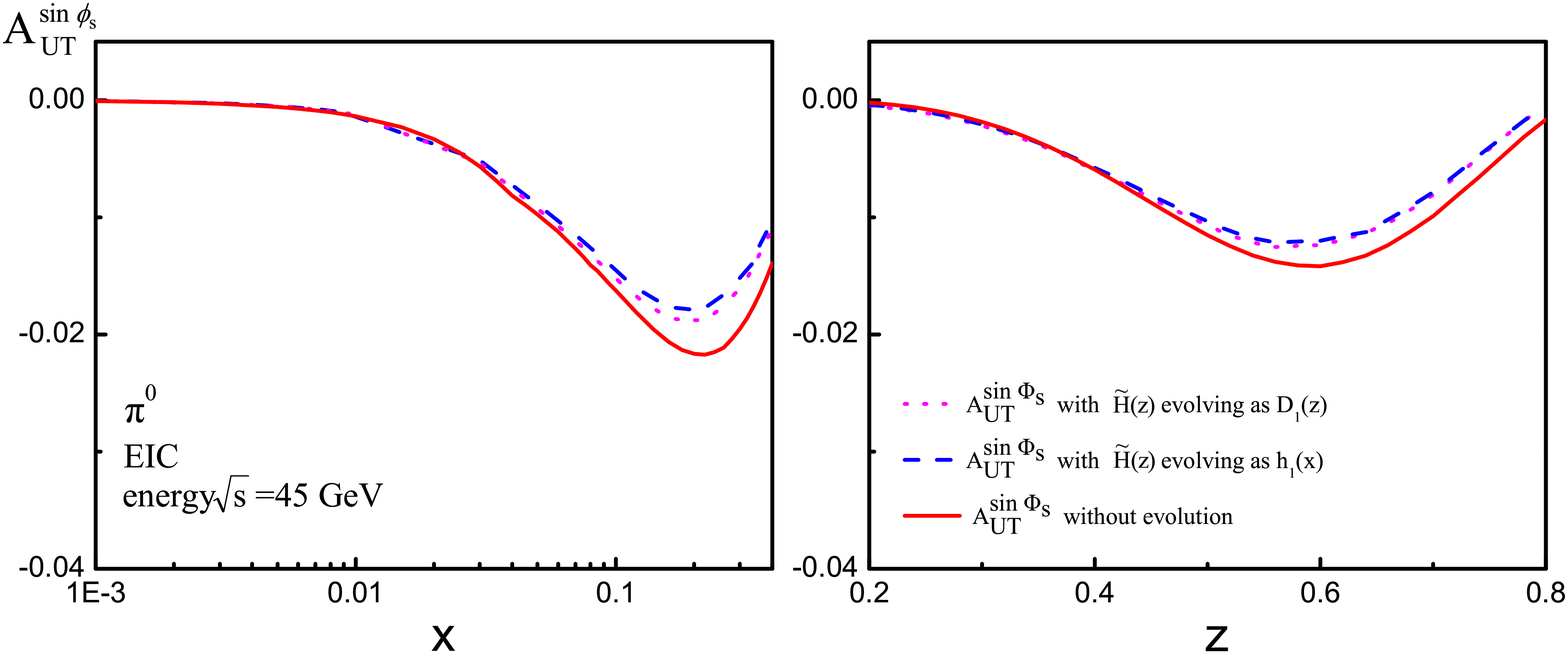}
\includegraphics[width=1.7\columnwidth]{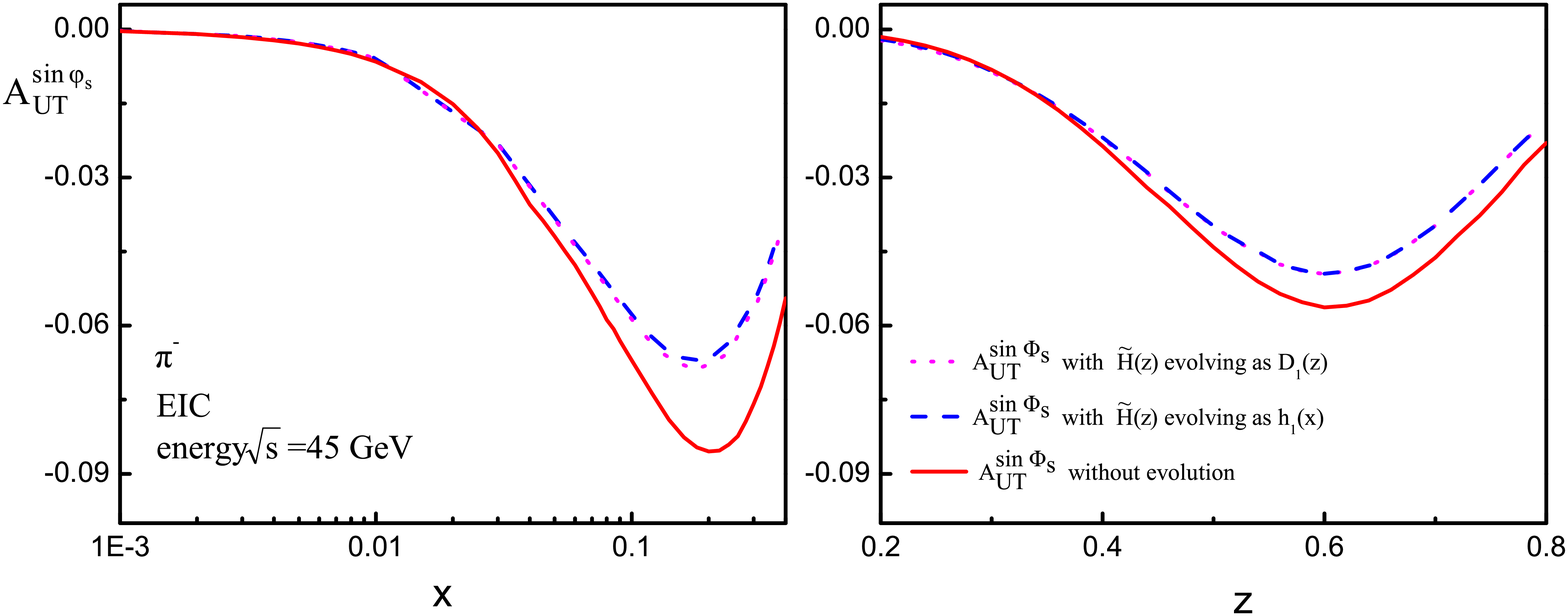}
\caption{Transverse SSA $A_{UT}^{\sin\phi_S}$ of $\pi^+$, $\pi^0$ and $\pi^-$ production in SIDIS at EIC for $\sqrt{s}=45$ GeV. The left panels show the $x$-dependent asymmetry, while the right ones for the $z$-dependent asymmetry.}
\label{fig:eic}
\end{figure*}

In this section, we apply the formalism introduced above to estimate the
$\sin \phi_S$ asymmetry at EIC.
As the kinematics at EIC covers a wide range of $Q$, it is necessary to consider the QCD evolution of the distribution and fragmentation functions, particularly, the transversity and the twist-3 function $\tilde{H}(z)$.
In the present calculation we implement their LO QCD evolution to study the corresponding impact on the $\sin\phi_S$ asymmetry.
To do this, we apply the HOPPET package~\cite{HOPPET} to perform the evolution, and we modify the HOPPET toolkit to include the LO DGLAP evolution kernel for the transversity distribution.
Since the QCD evolution of $\tilde{H}(z)$ is still unknown, in this calculation we adopt two different choices to evolve $\tilde{H}(z)$.
In the first choice we evolve $\tilde{H}(z)$ by employing the same evolution as the fragmentation function $D_1(z)$, following the scheme used in Ref.~\cite{1404.1033} for consistency.
As a comparison, in the second choice we assume that its evolution is the same as that of the transversity.
This is motivated by the fact that $\tilde{H}(z)$ is also a chiral-odd fragmentation function.
In the left and right panels of Fig.~\ref{fig:htilde}, we plot the $z$-dependence of the disfavored and favored twist-3 fragmentation functions $z\tilde H^{\pi^+/d}(z)$ and $z\tilde H^{\pi^+/u}$ at the scales $Q^2\ =1\,\mathrm{GeV}^2$ (shown by the solid lines) and $Q^2=100\,\textrm{GeV}^2$ (dotted lines: evolving the same as $D_1^q(z)$; dashed lines: evolving the same as $h_1^q(x)$),  respectively.
We find that in the region $z<0.4$, the evolution from low $Q$ to higher $Q$ increases the sizes of both $\tilde{H}^{\pi^+/u}$ and $\tilde{H}^{\pi^+/d}$; while it decreases their sizes in the region $z>0.4$.
In addition, there is slightly difference between the two evolved results of $\tilde{H}$ at the small-$z$ region.
However, if the uncertainties for the parametrization of $\tilde{H}$ were taken into account, the two curves calculated from two different evolution approaches would be statistically equivalent.

For the kinematical region that is available at EIC, we adopt the following choice~\cite{Accardi:2012qut}
\begin{align}
&Q^2>1 \mathrm{GeV}^2, \quad 0.001<x<0.4,\quad 0.01<y<0.95,\\\nonumber
&0.2<z<0.8,\quad \sqrt{s}=45\ \mathrm{GeV},\quad W>5\ \mathrm{GeV},
\end{align}
where $W$ is invariant mass of the virtual photon-nucleon system and $W^2=(P+q)^2\approx \frac{1-x}{x}Q^2$.
Using the above kinematical configuration and applying Eqs.~(\ref{eq:autx}) and (\ref{eq:autz}),
we predict the $\sin\phi_S$ asymmetry of pion production in SIDIS at EIC after the transverse momentum of pion is integrated.
The corresponding results are plotted in Fig.~\ref{fig:eic}, in which the upper, middle and lower panels show the $\sin \phi_S$ asymmetries at EIC for $\pi^+$, $\pi^0$ and $\pi^-$, respectively.
In each panel, we plot the asymmetries as functions of $x$ (left figure) and $z$ (right figure).
The solid lines denote the asymmetries without considering the evolution of the distribution and fragmentation functions in Eqs.~(\ref{eq:autx}) and (\ref{eq:autz}) (calculated at a fixed scale $Q^2=1\,\textrm{GeV}^2$). The dotted and dashed lines correspond to the two different treatments on the evolution of $\tilde{H}^q(z)$.

We find that the $\sin\phi_S$ asymmetries for the charged and neutral pion production are sizable, about several percent.
Therefore, there is a great opportunity to measure the $\sin\phi_S$ asymmetry in SIDIS at a future EIC.
In addition, the asymmetry for $\pi^+$ is positive, whereas the asymmetries for $\pi^-$ and $\pi^0$ are both negative;
and the asymmetry for $\pi^-$ is somewhat larger than those for $\pi^+$ and $\pi^0$.
This is because $h_1^u$ and $\tilde{H}^{\pi^+/d}$ are positive, while $h_1^d$ and $\tilde{H}^{\pi^+/u}$ are negative.
The numerical results show that the asymmetry for $\pi^0$ is close to $(A_{UT}^{\sin\phi_S\,,\pi^+}+A_{UT}^{\sin\phi_S\,,\pi^-})/2$, which is consistent with the assumption that $\tilde{H}^{\pi^0/q}$ is the average of $\tilde{H}^{\pi^+/q}$ and $\tilde{H}^{\pi^-/q}$.
We also find that the $x$-dependent asymmetry has a peak at the intermediate $x$ region, around $0.1<x<0.2$.
For both charged and neutral pion productions, the asymmetry reaches the maximum magnitude at the region $z\sim 0.6$.

An important observation is that the evolution effect for the $\sin\phi_S$ asymmetry is substantial in certain kinematical region at EIC.
First of all, as shown by the dotted lines ($\tilde{H}(z)$ evolves the same as $D_1(z)$) in Fig.~\ref{fig:eic}, the magnitudes of the $x$-dependent and $z$-dependent asymmetries for $\pi^+$, $\pi^0$ and $\pi^-$ are all reduced by QCD evolution.
Secondly, at small-$x$ region ($x<0.02$) the evolution does not affect the $x$-dependent asymmetries, while the evolution effect is sizable in the valence-$x$ region, especially in the case of $\pi^-$ production.
For the $z$-dependence asymmetries, the evolution effect may be observed in the region $z>0.4$.
Thirdly, the evolution effect for $\pi^-$ production is stronger than that for $\pi^+$ and $\pi^0$ production.
Finally, the uncertainties of the fragmentation function $\tilde{H}(z)$, which will result in the uncertainties of the asymmetries, are not considered in our calculation. 
If the uncertainties for the asymmetries were included, most likely the evolution effects from
the statistical viewpoint are not so dramatic in the kinematical region of EIC.
Nevertheless, the evolution almost does not change the signs and the shapes of the asymmetries.
As a comparison, we also show the asymmetries (the dashed lines) with $\tilde{H}$ evolving the same as $h_1$.
It is found that the asymmetries in this case are similar to the results calculated from the $D_1$ evolution for $\tilde{H}$, although slightly difference is observed in the $x$-dependent asymmetries for $\pi^0$ and $\pi^-$ production.

\section{Conclusion}
\label{conclusion}

In this work, we have implemented the twist-3 collinear fragmentation function $\tilde{H}(z)$ to study the $\sin\phi_S$ transverse SSA at EIC through the coupling $h_1(x)\otimes \tilde{H}(z)$, in the particular case that the transverse momentum of the final-state hadron is integrated out.
In our estimate we applied the standard parametrization for the transversity and the available extraction for the fragmentation function $\hat H_{FU}^{\Im}(z,z_1)$.
In addition, the LO evolution effects for the distribution and fragmentation functions were included.
The numerical prediction shows that the asymmetries for the charged and neutral pions are all sizable, about several percent.
Therefore, it is quite promising that the $\sin\phi_S$ asymmetries of meson production in SIDIS could be measured at the kinematics of EIC.
We also found the inclusion of evolution effect may be important for the interpretation of future experimental data.
In conclusion, our study demonstrates that it is feasible to access the transversity via transverse SSA of single meson production in SIDIS within the framework of collinear factorization.

\section*{Acknowledgements}
This work is partially supported by the National Natural Science
Foundation of China (Grants No.~11575043, and No.~11120101004), and
by the Qing Lan Project.

\end{document}